\documentstyle[12pt,preprint]{aastex}

\begin{document}

\def\wisk#1{\ifmmode{#1}\else{$#1$}\fi}

\def\lt     {\wisk{<}}
\def\gt     {\wisk{>}}
\def\le     {\wisk{_<\atop^=}}
\def\ge     {\wisk{_>\atop^=}}
\def\lsim   {\wisk{_<\atop^{\sim}}}
\def\gsim   {\wisk{_>\atop^{\sim}}}
\def\kms    {\wisk{{\rm ~km~s^{-1}}}}
\def\Lsun   {\wisk{{\rm L_\odot}}}
\def\Zsun   {\wisk{{\rm Z_\odot}}}
\def\Msun   {\wisk{{\rm M_\odot}}}
\def\um     {$\mu$m}
\def\mic     {\mu{\rm m}}
\def\sig    {\wisk{\sigma}}
\def\etal   {{\sl et~al.\ }}
\def\eg     {{\it e.g.\ }}
 \def\ie     {{\it i.e.\ }}
\def\bsl    {\wisk{\backslash}}
\def\by     {\wisk{\times}}
\def\half {\wisk{\frac{1}{2}}}
\def\third {\wisk{\frac{1}{3}}}
\def\nwm2sr {\wisk{\rm nW/m^2/sr\ }}
\def\nw2m4sr {\wisk{\rm nW^2/m^4/sr\ }}

\title{New measurements of cosmic infrared background fluctuations
from early epochs.}

\author{
A. Kashlinsky\altaffilmark{1,2,4}, R. G. Arendt\altaffilmark{1,2},
J. Mather \altaffilmark{1,3}, S. H. Moseley \altaffilmark{1,3} }
\altaffiltext{1}{Observational Cosmology Laboratory, Code 665,
Goddard Space Flight Center, Greenbelt MD 20771}
\altaffiltext{2}{SSAI} \altaffiltext{3}{NASA}
\altaffiltext{4}{e--mail: kashlinsky@stars.gsfc.nasa.gov}

\begin{abstract}
Cosmic infrared background fluctuations may contain measurable
contribution from objects inaccessible to current telescopic
studies, such as the first stars and other luminous objects in the
first Gyr of the Universe's evolution. In an attempt to uncover
this contribution we have analyzed the GOODS data obtained with
the Spitzer IRAC instrument, which are deeper and cover larger
scales than the Spitzer data we have previously analyzed. Here we
report these new measurements of the cosmic infrared background
(CIB) fluctuations remaining after removing cosmic sources to
fainter levels than before. The remaining anisotropies on scales
$\sim\!0.5^\prime\!-\!10^\prime$ have a significant clustering
component with a low shot-noise contribution. We show that these
fluctuations cannot be accounted for by instrumental effects, nor
by the Solar system and Galactic foreground emissions and must
arise from extragalactic sources.
\end{abstract}

\section{Introduction}

Cosmic infrared background (CIB) contains emissions from all
luminous including those inaccessible to current telescopic
studies (see Kashlinsky 2005 for recent review). If the first
stars (commonly called Population III) were massive
\cite{abell,bromm,brommlarson}, they could produce significant
levels of the CIB with measurable fluctuations
\cite{santos,sf2003,kagmm,cooray,madausilk,komatsu}. Recently, in
an attempt to uncover these signatures of early stars, we have
detected significant CIB fluctuations after removing galaxies to
faint limits in deep Spitzer data (Kashlinsky et al 2005;
hereafter KAMM). The measured fluctuations cannot arise from the
Solar System and Galactic foreground emissions, or from the
remaining galaxies and must come from early extragalactic sources,
particularly the first stars. To solidify these findings we have
now analyzed deeper datasets from the GOODS survey \cite{goods}
covering different and larger areas of the sky and with observing
procedures which allow us to test for a wide range of systematic
errors in the data.

In this {\it Letter} we present the new measurements using the
GOODS Spitzer/IRAC data at 3.6, 4.5, 5.8 and 8 \um.  At the
magnitude limit of the QSO1700 data we measure a similar CIB
signal consistent with its cosmological origin. After removing
still fainter sources, we find that the bulk of the CIB
fluctuations remain out to $\sim$10$^\prime$, the limit of our
fields. This shows that the signal originates in populations
significantly fainter than our cutoff. We also test for possible
systematic errors in the fluctuations by cross-correlating the
data for the same regions taken at different epochs and in
different detector orientations, showing only a small contribution
from the instrument systematics. The Galactic and zodiacal
foregrounds are generally also small, although as in the QSO1700
data, we find a possibility of a significant cirrus pollution of
the Channel 4 fluctuations. Cosmological interpretation of the
results is given in a separate paper (Kashlinsky, Arendt, Mather
\& Moseley 2006).

\section{Data processing}

Table 1 shows parameters of the data used in this and earlier
(KAMM) analyses. The new data from the GOODS Spitzer Legacy
Program
(PID=194, pipeline version=S11.4.0, except S11.0.2 for CDFS Epoch
1, Dickinson et al 2003) come from measurements in two parts of
the sky, HDFN in the North and CDFS in the South, observed at two
different Epochs, E1 and E2, $\sim$6 months apart. The two epochs
had different detector orientations rotated by $\sim$180$^\circ$
with respect to the sky, allowing a test for certain instrumental
effects and zodiacal light. The final mosaics were assembled from
the individual Basic Calibrated Data (BCD) frames which were
self-calibrated using the method of \cite{calibration} similarly
to KAMM; no extended source calibration factors were applied to
the data. To evaluate the random noise level of the maps,
alternate calibrated frames for each dataset were mapped into
separate ``A'' and ``B'' mosaics.

The assembled maps were cleaned of resolved sources in two steps:
First, an iterative procedure was applied which computes the
standard deviation, $\sigma$, of the image, and masks pixels
exceeding $N_{\rm cut} \sigma$ along with $N_{\rm
mask}\!\times\!N_{\rm mask}$ surrounding pixels until no pixels
exceed  $N_{\rm cut} \sigma$. We adopted $N_{\rm mask}$=3 and
$N_{\rm cut}$=4, when enough pixels ($\geq$65\%) remain for robust
Fourier analysis. Second, similar to KAMM, we subtracted the Model
of the individual sources using a variant of the CLEAN algorithm
\cite{clean}. In a normal CLEAN procedure, the clean components
are convolved with an idealized clean beam to produce a map
without sidelobe artifacts; we stop short of this, working with
the residual map from which the dirty beam and its artifacts have
been removed. We construct the Model from the original unmasked
mosaics for each channel as follows: (1) the maximum pixel
intensity is located, (2) the PSF is then scaled to half of this
intensity and subtracted from the image, (3) the process is
iterated, saving intermediate results.

Because spectroscopic redshifts are unavailable and resolving
individual sources is difficult in the confusion limit of these
observations, we removed the sources via the Model to a fixed
level of the shot noise. Its amplitude due to remaining sources is
$P_{\rm SN} = \int_{m_{\rm lim}}^\infty [S(m)]^2 dN/dm \; dm$.
$S(m)$ is the flux of magnitude $m$ in nW/m$^2$; diffuse flux is
defined as $\nu I_\nu$, with $I_\nu$ being the surface brightness
in MJy/sr. With the Model iteration where the maps remain close to
a fixed $P_{\rm SN}$ we can probe the CIB fluctuations produced by
populations below a relatively well-defined flux threshold. In
principle, the remaining instrument noise, truncated of its high
{\it positive} peaks, may have an imprint of the beam and mimic
part (or even most) of the shot noise from cosmic sources. In that
case, the shot noise we present in Fig. 1 represents an {\it
upper} limit on the contribution from cosmic sources making our
conclusions below stronger.

The maps, clipped and with the Model subtracted, were
Fourier-transformed and subjected to all the same tests as in
KAMM; the additional tests possible with this data are described
below. The fluctuation field, $\delta F(${\mbox{\boldmath$x$}),
was not weighted for the results presented here (although we
checked that weighting by the observation time in each pixel does
not lead to any appreciable changes), and its Fourier transform,
$f($\mbox{\boldmath$q$}$)= \int \delta F($\mbox{\boldmath$x$}$)
\exp(-i$\mbox{\boldmath$x$}$\cdot$\mbox{\boldmath$q$}$) d^2x$ was
calculated using the fast Fourier transform. The 2-D power
spectrum is defined as $P_2(q)\equiv \langle |f_q|^2\rangle$ and
in this definition a typical flux fluctuation is $\simeq
\sqrt{q^2P_2(q)/2\pi}$ on the angular scale of wavelength $2
\pi/q$. The blanked pixels were assigned $\delta F$=0, thereby not
adding power to the eventual power spectrum. Muxbleed was removed
by zeroing the corresponding frequencies in the $(u,v)$ plane
before computing the power spectrum of the maps, $P_S$. We
subtracted remaining linear gradients from the maps before
clipping and again after clipping. The noise power spectrum,
$P_N$, was evaluated from the $\frac{1}{2}(A-B)$ data using the
mask from the main maps. The remaining power spectrum was
evaluated as $P=P_S - P_N$.

\section{Results}

Fig. \ref{fig:results} shows results from the new data compared to
the earlier QSO1700 data analysis. As discussed above, in the
present analysis we can regulate the faint source removal from the
maps by fixing the floor level of the shot noise, whose amplitude
$P_{\rm SN}$ was evaluated from the fits to the small scale power
spectrum using our model for the beam. Because we mapped the GOODS
data onto 1.2$^{\prime\prime}$ pixels (instead of
$0.6^{\prime\prime}$ for the QSO1700 data) the beam used for
generating the Model was a slightly smoothed version of that used
in the QSO1700 data.

The upper panels in the figure show the CIB fluctuations in the
GOODS fields evaluated at the Model iteration when $P_{\rm SN}$
roughly corresponds to that in the earlier QSO1700 analysis. The
figure shows CIB fluctuations consistent with those in the earlier
analyzed QSO1700 dataset. Our results are consistent that at the
same level of $P_{\rm SN}$ all data are probing the same
populations. (At the deeper Model iterations extra galaxy
populations were removed in the present analysis compared to the
QSO1700 field and at shallower Model iterations extra galaxy
populations were removed in the QSO1700 field). Because of longer
integration in the new data, we can remove sources to still lower
$P_{\rm SN}$. The lower panels of Fig. \ref{fig:results} show the
CIB fluctuations remaining in the maps at the lowest common value
of $P_{\rm SN}$; the signal shown here comes from still fainter
sources than the limits in the earlier KAMM analysis.

\section{Discussion}

The following can, in principle, contribute to the detected
fluctuations 1) instrumental (systematic and random), 2) source
artefacts, 3) Solar System and Galactic foregrounds, and 4)
extragactic sources. We briefly discuss the contributions of each
and conclude that (with the possible exception of the 8\um\ data)
the detected fluctuations are due to CIB from extragalactic
sources below our removal threshold set by the shot-noise
amplitude.

{\it Checking for systematics: E1 vs E2}. In measuring the faint,
low spatial frequency backgrounds, control of systematic errors
remains a major challenge.  Of particular concern is scattered
light  in the Spitzer and IRAC optical systems, which can spread
the light from point sources over large spatial scales.  Some
scattered light retains a fixed relative position with respect to
the originating point source. Other scattered light gets to the
focal plane after multiple reflections, and its spacing with
respect to the originating source will vary as the source is moved
in the field of view. Finally, some scattered light may arise from
sources outside the field of view, and may change its illumination
of the detector in a complex way as the telescope is moved on the
sky. Given that scattered light and  detector artifacts can cause
structure in the image, it is important that we use observations
for these studies which allow us to  evaluate the size of such
possible effects in the images we analyze.

To search for  instrumental sources of large scale power, we used
the partially overlapping E1 and E2 data of the HDN and CDFS
regions in which the telescope and optical system are rotated by
$\sim$180 degrees with respect to the field of view. Using such
observations, we can compare the source-removed sky maps from the
two epochs to separate structure which is unchanged in inertial
coordinates, and thus presumably arising from the sky, and that
which changes, which could represent an instrumental contribution.

We selected overlapping subfields for the final mosaicked images
which had fairly homogeneous exposures. The latter limited the
selected areas to approximately the same size: $142\times504$
arcsec in the CDFS field and $149\times 504$ arcsec in the HDFN
area. We then computed the subfield and cross-subfield correlation
functions and coefficients to verify that the signal is the same
at each Epoch and is detector-orientation independent. Having
carried out the point source subtraction and gradient removal on
the two epochs, we  compared the correlation function of the
difference map to that of each of the individual maps, and also
did a cross correlation analysis. In both cases, we find that all
of the large scale power arises from the sky down to the presented
statistical error bars; this rules out any significant
contribution from instrumental scattered light and detector
artifacts.

We also estimated that there is at most only a small contribution
to the measured power spectrum from image artefacts, such as e.g.
associated with occasional strong muxbleed. This was done by
selecting sub-regions excluding the areas of obvious artefacts.
(This is responsible for a slight excess in power at 3.6 \um\
around 1-2 arcmin for one of the fields).

{\it Procedural sources of large-scale power}. Because IRAC's
calibration does not include zero-flux closed-shutter data, a
degeneracy remains in the absolute zero point of the solution. A
similar degeneracy is present with respect to first-order
gradients in the mosaics. However higher-order gradients are not
degenerate. Because of these degeneracies, linear gradients have
been fit and subtracted from the self-calibrated mosaicked images.
The zero-level is unimportant to this study, so all analyzed maps
are set to a mean intensity equal to zero.

We find - via simulations - that in small, but non-negligible
number of cases the gradient-subtraction can also remove the
genuine cosmic power at the largest scales and the power shown
there should thus be treated as the {\it lower} limit on any
cosmic fluctuations. However, we find in the QSO1700, the HDFN-E1
and HDFN-E2 analyses that the results do not change appreciably
even if the extra-subtraction is done at {\it each} step of the
Model iteration. The CDFS fields, which have a common overlap at
both Epochs, exhibit more sensitivity (at the level of $\sim 20\%$
in the largest bins) to this extra gradient subtraction. Taken
together this is consistent with the cosmic nature of the
fluctuations shown in Fig. \ref{fig:results} and the CDFS field
happening to be statistically more sensitive to the extra gradient
subtraction.

The  possible contributions from incompletely removed galaxies are
also small as discussed in KAMM. Briefly: 1) the measured
fluctuations are independent of the masking and clipping
parameters. E.g. when clipping is $N_{\rm cut}$=2 only 6\% of the
maps remained \cite{kamm}, but the correlation function, which
replaces the power spectrum as a measure of large-scale
correlations for such deeply cut maps, remained practically the
same. 2) The results are invariant when the masking size around
each clipped pixel is increased (from $N_{\rm mask}$=3 to 7). This
mask is larger than the typical galaxy size at $z\gsim$0.1 so
individual galaxies are expected to have been removed completely.
The Model further removes much of the remaining emissions at
larger angles. 3) We measure the clustering component from $\sim
0.5^\prime$ to $\sim$5-10$^\prime$, which subtend 1 to 10 Mpc at
$z$=0.1,1. If the power comes from the incompletely removed local
galaxies, its angular spectrum should reflect the slope of the
observed galaxy two-point correlation function, contrary to Fig.
\ref{fig:results}. 4) The results are the same for all fields at
the fixed level of shot-noise. The level of the remaining $P_{\rm
SN}$ fixes the amount of the remaining flux from incompletely
removed sources of full magnitude $m$. The contribution from them
to the large scale clustering should then have been a proportional
fraction of the CIB clustering produced by the parental sources.
5) By construction in their results, KAMM identified the
appropriate Model iteration number where the clean components
become largely uncorrelated with the sources in the original
image. (Here we use $P_{\rm SN}$ as the alternative criterion).

{\it Zodiacal and Galactic foregrounds}. The cirrus flux in Table
1 are the SPOT estimates of ISM intensity, based on the Schlegel
et al (1998) 100 \um\ IRAS intensity and temperature maps and a
spectral scaling relation derived from DIRBE, ISO and AROME
observations of the ISM
(http://ssc.spitzer.caltech.edu/documents/background/bgdoc\_release.html;
Reach \& Boulanger 1997). Fluctuations in the cirrus emission are
assumed to be at the 1\% level as suggested by the intensity of
the fluctuation spectrum of a low galactic latitude field in which
cirrus emission was clearly evident at 8 \um\ (KAMM). As in the
QSO1700 data, the cirrus may contribute a non-negligible
contribution at 8 \um, and prevents us from isolating the
cosmological signal there. Even assuming that the entire signal at
8 \um\ is produced by cirrus and taking the mean Galactic cirrus
energy distribution \cite{cirrus} gives an upper limit on the
cirrus emission at shorter wavelengths well below the signal we
detect.

The range in zodiacal light intensity in the time spanned by the
observations is shown in Table 1. Its fluctuations are a factor of
$\simeq$4-15 below those from cirrus. A limit of $\delta F <$0.1
\nwm2sr at 8 \um\ due to zodiacal light was derived by KAMM from
the observed dispersion of the data at two epochs separated by
about six months. This limit at 8 \um\ (with relative fluctuations
at a much lower level than the $<$0.2\% limit by Abraham et al.
(1997) at 25 \um) was then scaled to the shorter wavelengths using
the zodiacal light spectrum derived from COBE/DIRBE data (Kelsall
et al. 1998). In the present study too we find no evidence for
zodiacal emission above the instrument noise levels; the arcminute
limits on zodiacal fluctuations from subtracting the two Epochs
are $\delta F \lsim$0.05 \nwm2sr at 8 \um.

{\it Extragalactic sources}. The detected signal is thus due to
CIB fluctuations from extragalactic sources, such as ordinary
galaxies and the putative Population III. KAMM estimated that the
CIB flux from the remaining galaxies was only $\simeq$0.15 \nwm2sr
so that they were unlikely to account for the strong clustering
signal. The present data enable us to eliminate intervening
sources down to lower levels of the shot noise, or lower flux
limits. The detected signal has to originate in still fainter
sources. In a companion paper (Kashlinsky, Arendt, Mather \&
Moseley 2006) we discuss the constraints our results (both the
shot-noise and clustering components of the fluctuations) place on
the nature of the sources contributing them. We show that the
signal at 3.6, 4.5 \um\ must arise from very faint populations
with individual fluxes $\lsim$10-20 nJy and that the amplitude of
the fluctuations at arcminute scales requires these populations to
be significantly more luminous per unit mass than the present-day
ones.

This work is supported by NSF AST-0406587 and NASA Spitzer
NM0710076 grants.


\clearpage

\begin{deluxetable}{lcccccc}
\tabletypesize{\scriptsize} \tablewidth{0pt}
\tablecaption{Analyzed Fields\label{table1}} \tablehead{
\colhead{Parameter} & \colhead{Channel} &\colhead{QSO1700} &
\colhead{HDFN-E} & \colhead{HDFN-E2} & \colhead{CDFS-E1} &
\colhead{CDFS-E2} } \startdata
($l_{\rm Gal},b_{\rm Gal}$) (deg) & \nodata &  ($94.4,36.1$) & ($125.9,54.8$) & ($125.9,54.8$) & ($223.6,-54.4$) & ($223.6,-54.4$) \\
($\lambda_{\rm Ecl},\beta_{\rm Ecl}$) (deg) & \nodata & ($194.3, 83.5$) & ($148.4, 57.3$) & ($148.4, 57.3$) & ($41.1, -45.2$) & ($41.1, -45.2$) \\
Size (arcmin) & \nodata & $5.1\!\times\!11.5$ & $10.2\!\times\!10.2$ & $10.2\!\times\!10.2$ & $8.8\!\times\!8.4$\tablenotemark{a} & $9.0\!\times\!8.4$\tablenotemark{b}\\
$\langle t_{\rm obs}\rangle$ (hr)\tablenotemark{c} & \nodata & 7.8 & 20.9 & 20.7 & 23.7 & 22.4 \\
$P_{\rm SN}$ $({\rm nW}2{\rm m}^{-4}{\rm sr})$ & 1 (3.6 $\micron$) & $5.8\!\times\!10^{-11}$ & $1.9\!\times\!10^{-11}$ & $1.9\!\times\!10^{-11}$ & $2.2\!\times\!10^{-11}$ & $2.3\!\times\!10^{-11}$\\
Flux: zodi (nW/m$^2$/sr) & 1 (3.6 $\micron$) & 32 & 45--47 & 37--38 & 48--51 & 48--49\\
Flux: cirrus (nW/m$^2$/sr) & 1 (3.6 $\micron$) & 2.5 & 0.8 & 0.8 & 0.8 & 0.8\\
$P_{\rm SN}$ $({\rm nW}2{\rm m}^{-4}{\rm sr})$ & 2 (4.5 $\micron$) & $6.0\!\times\!10^{-11}$ & $1.1\!\times\!10^{-11}$ & $1.0\!\times\!10^{-11}$ & $9.5\!\times\!10^{-12}$ & $1.1\!\times\!10^{-11}$\\
Flux: zodi (nW/m$^2$/sr) & 2 (4.5 $\micron$) & 132 & 174--185 & 146--153 & 189--204 & 165--176\\
Flux: cirrus (nW/m$^2$/sr) & 2 (4.5 $\micron$) & 2.7 & 1.3 & 1.3 & 1 & 1 \\
$P_{\rm SN}$ $({\rm nW}2{\rm m}^{-4}{\rm sr})$ & 3 (5.8 $\micron$) & $6.0\!\times\!10^{-10}$ & $1.2\!\times\!10^{-10}$ & $1.3\!\times\!10^{-10}$ & $1.5\!\times\!10^{-10}$ & $1.5\!\times\!10^{-10}$\\
Flux: zodi (nW/m$^2$/sr) & 3 (5.8 $\micron$) & 873 & 1192--1250 & 996--1026 & 1256--1327 & 1179--1243\\
Flux: cirrus (nW/m$^2$/sr) & 3 (5.8 $\micron$) & 7.8 & 36. & 3.6 & 2.6 & 2.6 \\
$P_{\rm SN}$ $({\rm nW}2{\rm m}^{-4}{\rm sr})$ & 4 (8 $\micron$) & $4.4\!\times\!10^{-10}$ & $1.0\!\times\!10^{-10}$ & $1.1\!\times\!10^{-10}$ & $1.1\!\times\!10^{-10}$ & $1.4\!\times\!10^{-10}$\\
Flux: zodi (nW/m$^2$/sr) & 4 (8 $\micron$) & 1723 & 2411--2500 & 2012--2054 & 2509--2614 & 2454--2562\\
Flux: cirrus (nW/m$^2$/sr) & 4 (8 $\micron$) & 33.8 & 16.1 & 16.1 & 10.1 & 10.1\\
\enddata
\tablenotetext{a}{$9.0\!\times\!8.4$ arcmin for Channels 2 and 4.}
\tablenotetext{b}{$8.8\!\times\!8.4$ arcmin for Channels 2 and 4.}
\tablenotetext{c}{For Channel 1. Other channels may vary by as
much as 0.5 hr, except for the QSO1700 region where $\langle
t_{\rm obs}\rangle = 9.2$ hr.}
\end{deluxetable}
\clearpage

\begin{figure}
\plotone{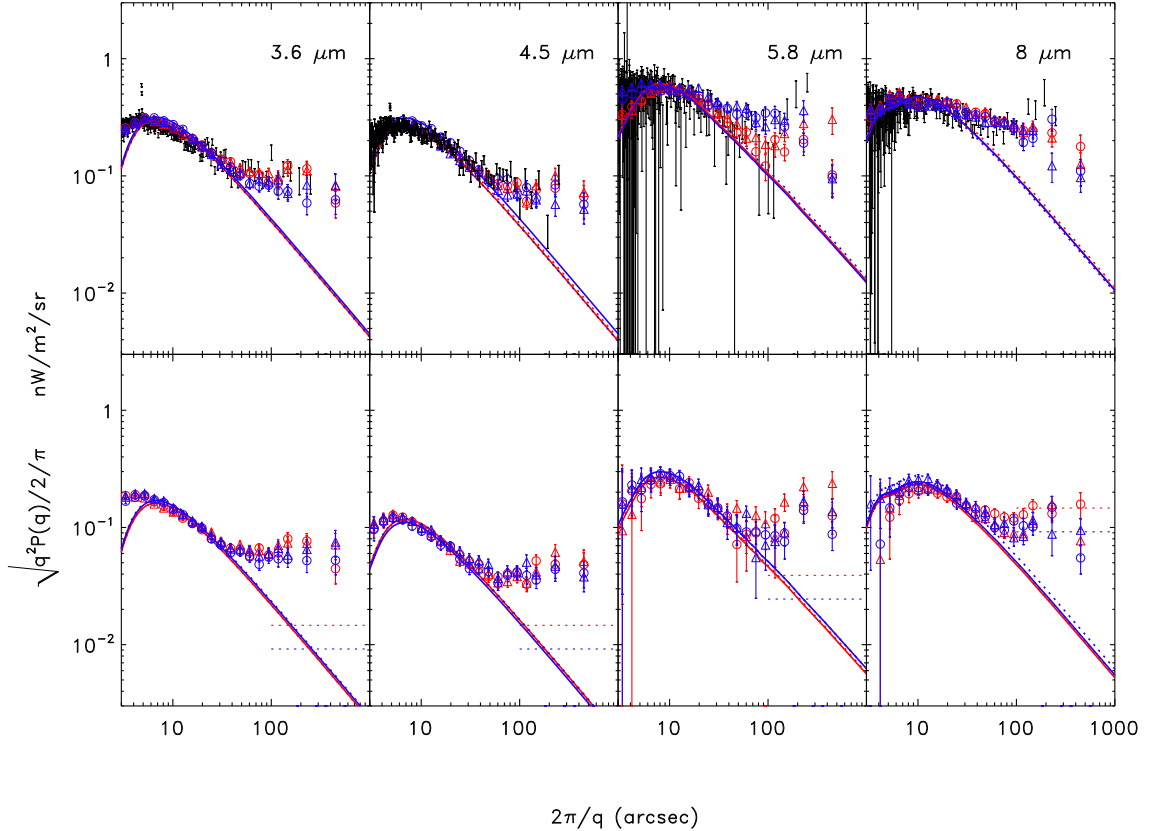} \caption{CIB fluctuations from the analyzed
fields. The fluctuation spectrum was evaluated by averaging over
the concentric rings center at given $q$ and the errors shown
correspond to the cosmic variance for the estimates, i.e. the
Poissonian errors of relative amplitude $N_q^{-1/2}$ with $N_q$
being the number of independent Fourier elements used in
determining the power at $q$. Top: the CIB fluctuations were
evaluated for the shot-noise levels approximately corresponding to
the $P_{\rm SN}$ of the QSO1700 data. Black error bars show the
QSO1700 fields results from \cite{kamm}. Red symbols correspond to
the HDFN fields and blue symbols denote the CDFS data results.
Triangles correspond to E1 and circles to E2. The lines show the
fluctuations due to shot-noise from the remaining sources. Bottom:
Same as top shown for the lowest levels of $P_{\rm SN}$ reached
with the new data. The blue and red dotted lines show the
estimated cirrus fluctuations for the CDFS and HDFN fields
respectively. } \label{fig:results}
\end{figure}


\begin{thebibliography}{3}
\bibitem [Abell 2002]{abell}{Abell, T. 2002, Science, 295, 93}
\bibitem [\'Abrah\'am et al 1997]{abraham}{\'Abrah\'am, P., Leinert, Ch., \& Lemke, D. 1997, A\&A, 328, 702}
\bibitem [Arendt et al 1998]{cirrus}{Arendt, R. et al 1998, Ap.J., 508,74}
\bibitem [Bromm et al 1999]{bromm}{Bromm, V. et al 1999, Ap.J., 527,
L5}
\bibitem [Bromm \& Larson 2004]{brommlarson}{Bromm, V. \& Larson,
R. 2004, Ann. Rev. A. A., 42, 79}
\bibitem [Cooray et al 2004]{cooray}{Cooray, A. et al 2004, Ap.J., 606, 611}
\bibitem [Dickinson et al 2003]{goods}{Dickinson, M. et al 2003, ``The great observatories origins deeps survey",
in ``The mass of galaxies at low and high redshift", ed. R. Bender
\& A. Renzini, astro-ph/0204213}
\bibitem [Fernandez \& Komatsu 2006]{komatsu}{Fernandez, E. R. \&
Komatsu, E. 2005, Ap.J., 646, 703}
\bibitem [Fixsen et al 2000]{calibration}{Fixsen, D. J.,
Moseley, S. H. \& Arendt, R. G. 2000, Ap. J. Suppl., 128, 651}
\bibitem [H$\ddot{\rm o}$gbom 1974]{clean}{H$\ddot{\rm o}$gbom, J. 1974, Ap.J.Suppl., 15,417}
\bibitem [Kashlinsky 2005]{review}{Kashlinsky, A. 2005, Phys. Rep., 409,
361-438}
\bibitem [Kashlinsky et al 2004]{kagmm}{Kashlinsky, A., Arendt, R., Gardner, J.P., Mather, J.C., \& Moseley, S.H. 2004, Ap.J., 608, 1 }
\bibitem [KAMM]{kamm}{Kashlinsky, A., Arendt,
R., Mather, J.C. \& Moseley, S.H. 2005, Nature, 438, 45}
\bibitem [Kashlinsky et al 2006]{interpretation}{Kashlinsky, A., Arendt,
R., Mather, J.C. \& Moseley, S.H. 2006, Ap.J., submitted.}
\bibitem [Kelsall et al 1998]{kelsall}{Kelsall, T. et al 1998,
Ap.J., 508, 44}
\bibitem [Madau \& Silk 2005]{madausilk}{Madau, P. \& Silk, J.
2005,MNRAS, 359, L37}
\bibitem [Reach \& Boulagger 1997]{reach}{Reach, W. T. \& Boulanger, F. 1997, in
``Infrared emission from interstellar dust in the local
interstellar medium: The Local Bubble and Beyond", eds. D.
Breitschwerdt, M. J. Freyburg, \& J. Trumper;352-362,
Springer:Berlin}
\bibitem [Santos et al 2002]{santos}{Santos, M.R., Bromm, V., Kamionkowski, M. 2002,MNRAS,336,1082}
\bibitem [Salvaterra \& Ferrara 2003]{sf2003}{Salvaterra, R. \& Ferrara, A. 2003, MNRAS, 339, 973}
\bibitem[Schlegel et al 1998]{schlegel}{Schlegel, D., Finkbeiner,
D. \& Davis, M. 1998, Ap.J., 500, 525}

\end{thebibliography}
\end{document}